\documentclass[twocolumn,showpacs,preprintnumbers,amsmath,amssymb]{revtex4}
\usepackage{graphicx}
\usepackage{dcolumn}
\usepackage{bm}

\begin{document}


\title{Detecting photon-photon scattering in vacuum at exawatt lasers}

\author{Daniele Tommasini,$^{1}$ Albert Ferrando,$^{2}$ Humberto Michinel,$^{1}$
Marcos Seco$^{3}$}

\affiliation{$^{1}$Departamento de F\'\i sica Aplicada. Universidade de Vigo.
As Lagoas E-32004 Ourense, Spain.}

\affiliation{$^{2}$Interdisciplinary Modeling Group, InterTech. Departament d'\`{O}ptica,
Universitat de Val\`{e}ncia. Dr. Moliner, 50. E-46100 Burjassot (Val\`{e}ncia),
Spain.}

\affiliation{$^{3}$Departamento de F\'{\i}sica de Part\'{\i}culas. Universidade de Santiago de Compostela.
15706 Santiago de Compostela, Spain.}


\begin{abstract}
In a recent paper, we have shown that the QED nonlinear corrections imply a
phase correction to the linear evolution of crossing electromagnetic waves in vacuum.
Here, we provide a more complete analysis,
including a full numerical solution of
the QED nonlinear wave equations for short-distance
propagation in a symmetric configuration.
The excellent agreement of such a solution with the result
that we obtain using our perturbatively-motivated
Variational Approach is then used to justify an
analytical approximation that can be applied in a more general case.
This allows us to find the most promising configuration
for the search of photon-photon scattering in optics experiments.
In particular, we show that our previous requirement of phase coherence between
the two crossing beams can be released.
We then propose a very simple experiment that can be performed at
future exawatt laser facilities, such as ELI,
by bombarding a low power laser beam with the exawatt bump.
\end{abstract}


\maketitle

\section{Introduction}

Radiative corrections in Quantum Electrodynamics (QED)
have been studied for 70 years, both theoretically and
experimentally\cite{Weinberg}.
Nevertheless, in the last decade they have
gained an increasing interest, following the extraordinary
advancements in the fields of Quantum and Nonlinear Optics.
On one hand, it has been noted that
a fundamental uncertainty in the number of photons
is unavoidably generated by QED radiative 
corrections\cite{tommasini0203},
eventually competing with the experimental errors.
On the other hand, the interchange of virtual
electron-positron pairs can produce classically forbidden
processes such as photon-photon scattering in vacuum.

Although it is a firm prediction of QED,
photon-photon scattering in vacuum
has not yet been detected, not even indirectly.
The rush for its discovery is then wide open.
A {\it hammer} strategy will be to build a photon-photon
collider\cite{phcollider}, based on an electron laser
producing two beams of photons in the $MeV$ range
(i.e., having wavelengths in the range of a few $fm$).
This will maximize the cross section for the process.
A second approach will be to perform experiments using ultrahigh power optical lasers,
such as those that will be available in the near future\cite{mourou06},
in such a way that the high density of
photons will compensate the smallness of the cross section.
In this case, the photon energies are well below the electron
rest energy, and the effect of photon-photon
collisions due to the interchange of virtual
electron-positron pairs can be expressed in terms of
the effective Euler-Heisenberg nonlinear Lagrangian\cite{Halpern,Heisenberg}.
This modifies Maxwell's equations for the average values of the electromagnetic
quantum fields\cite{mckenna} and affects the properties of the QED vacuum\cite{klein}.

Ultra intense photon sources are available thanks to the
discovery of chirped pulse amplification (CPA)\cite{strickland85} in the
late 80's and optical parametric chirped pulse amplification (OPCPA)
\cite{dubietis92} in the 90's. These techniques opened the door to a
field of research in the boundary between optics and experimental
high-energy physics, where lots of novelties are expected to
come in the next years. In fact, several recent works propose different configurations
that can be used to test the nonlinear optical response of the vacuum, e.g.
using harmonic generation in an inhomogeneous magnetic field\cite{ding92}, QED
four-wave mixing\cite{4wm}, resonant interactions in microwave cavities\cite{brodin},
or QED vacuum birefringence\cite{Alexandrov} which can be probed by x-ray pulses\cite{xray},
among others\cite{others}.

In a recent Letter\cite{prl99}, we have shown that photon-photon scattering in vacuum
implies a phase correction to the linear evolution of
crossing electromagnetic waves,
and we have suggested an experiment for measuring this effect
in projected high-power laser facilities like the European Extreme Light Infrastructure
(ELI) project\cite{ELI} for near IR radiation.
Here, we provide a more complete analysis,
including a full numerical solution of
the QED nonlinear wave equations for short-distance
propagation in a symmetric configuration.
The excellent agreement of such a solution with the result
that we obtain using a perturbatively-motivated
Variational Approach is then used to justify an
approximation that can be applied in a more general case.
This allows us to find the most promising configuration
for the search of photon-photon scattering in optics experiments.
In particular, we show that our previous
requirement of phase coherence between
the two crossing beams can be released.
We then propose a very simple experiment that can be performed at
future exawatt laser facilities, such as ELI,
by bombarding a low power laser beam with the exawatt bump.
The effect of photon-photon scattering will be detected by measuring
the phase shift of the low power laser beam, e.g. by
comparing it with a third low power laser beam.
This configuration is simpler, and significantly more sensitive,
than the one proposed in our previous paper.
Even in the first step of ELI, we find that the
resulting phase shift will be at least
$\Delta\Phi\approx 2\times 10^{-7} rad$,
which can be easily measured
with present technology.
Finally, we discuss how the
experimental parameters can be adjusted
to further improve the sensitivity.

The paper is organized as follows:

In Section II, following Ref. \cite{prl99},
we present the nonlinear equations
that replace the linear wave equation
when the QED vacuum effect is taken into account.

In Section III, we consider a linearly polarized wave,
describing the scattering of two counterpropagating plane waves
that have the same intensity and initial phase at the origin
(symmetric configuration).
In this case, we perform a numerical simulation
for short-distance propagation, which provides the first
known solution to the full QED nonlinear wave equation.

In Section IV, we study the same configuration as in Sect. III
by using a Variational Approximation. This allows us to obtain
an {\it analytical}
solution which is valid even for a long evolution.
After showing that this variational solution is
in good agreement with the numerical simulation of Sect. III,
we give a method to study different configurations that cannot
be integrated numerically in a simple and direct way.

In Section V, we apply our Variational Approximation to
the case of the scattering of a relatively low power
wave with a counterpropagating high power wave,
both traveling along the $z$ axis.
We allow an arbitrary initial phase relation between the two waves,
and obtain an analytical solution showing a phase shift of the
low power wave.

In Section VI, we discuss the possibility of detecting
photon scattering by measuring the phase shift
of two counterpropagating waves at ELI. We show that an
asymmetric configuration, in which only one high power
laser beam is used and the phase is measured on the low power
laser beam, is not only simpler to realize, but it also gives
a better sensitivity than the symmetric configuration.
The resulting proposed experiment will then allow to
detect photon-photon scattering as originated by QED theory.

In Section VII, we resume our conclusions, and discuss why
we think that our proposed experiment will be the simplest
and most promising way to detect photon-photon scattering
with optical measurements in the near future.

\section{The nonlinear equation for linearly polarized waves in vacuum}

In this section we introduce the equations that govern the evolution of 
the electromagnetic fields {\bf E } and {\bf B} when the 
QED effects are taken into account.
We assume that the photon energy is well below the
threshold for the production of electron-positron pairs,
$2 m_e c^2\simeq 1 MeV$. This means that we will only
consider radiation of wavelengths $\lambda\gg 2\times 10^{-13} m$,
which is always the case in optical experiments.
In this case, the QED effects can be described by the
Euler-Heisemberg effective Lagrangian density\cite{Heisenberg},

\begin{equation}
{\cal L}={\cal L}_0+\xi{\cal L}_Q={\cal L}_0
+\xi\left[{\cal L}_0^2+ \frac{7\epsilon_0^2 c^2}{4}({\bf E}\cdot {\bf B})^2\right],
\label{L}
\end{equation}
being
\begin{equation}
{\cal L}_0=\frac{\epsilon_0}{2}\left({\bf E}^2-{c^2\bf B}^2\right)
\label{L0}
\end{equation}
the linear Lagrangian density and $\epsilon_0$ and $c$ the dielectric
constant and the speed of light in vacuum, respectively. As it can be
appreciated in Eq. (\ref{L}), QED corrections are introduced by the parameter
\begin{equation}
\xi=\frac{8 \alpha^2 \hbar^3}{45 m_e^4 c^5}\simeq 6.7\times
10^{-30}\frac{m^3}{J}.
\label{constant_xi}
\end{equation}

This quantity has dimensions of the inverse of an energy density. This means that
significant changes with respect to linear propagation can be expected for values around
$\vert\xi{\cal L}_0\vert\sim 1$, corresponding to beam fluxes with electromagnetic energy densities given by
the time-time component of the energy-momentum tensor
\begin{equation}
T_{00}=\frac{\partial{\cal L}}{\partial (\partial_tA)}\partial_t A - {\cal L}
\gtrsim 2/\xi\simeq 3\times 10^{29} J/m^3.
\label{T00}
\end{equation}
While such intensities may have an astrophysical or cosmological importance,
they are not achievable in the laboratory. The best high-power
lasers that are being projected for the next few decades will be
several orders of magnitude weaker, eventually reaching energy
densities of the order $\rho\sim 10^{23} J/m^3$\cite{mourou06}. Therefore, we will
study here the ``perturbative'' regime, in which the non-linear
correction is very small, $\vert\xi{\cal L}_0\vert \ll 1$. As we
shall see, even in this case measurable effects can be
accumulated in the phase of beams of wavelength $\lambda$ traveling over
a distance of the order $\lambda \vert\xi{\cal L}_0\vert^{-1}$. Thus, current sensitive
techniques could be used to detect traces of QED vacuum nonlinearities.

Once the electromagnetic fields are expressed in terms of the four-component gauge
field $A^\mu=(A^0,{\bf A})$ as ${\bf B}=\nabla \wedge {\bf A}$ and
${\bf E}=-c \nabla A^0-\frac{\partial {\bf A}}{\partial t}$, the equations
of motion are given by the Variational Principle:

\begin{equation}
\frac{\delta \Gamma}{\delta A^\mu}=0,
\label{var_princ_gen}
\end{equation}
where $\Gamma\equiv \int {\cal L} d^4x$ is the QED effective
action. Instead of studying the resulting equations for the fields
${\bf E}$ and ${\bf B}$, that can be found in the literature\cite{mckenna,variational},
for the present purposes it is more convenient to consider the equations for the gauge field components $A^\mu$.
In general, these four equations cannot be disentangled. However, after some straightforward algebra
it can be seen that they admit solutions in the form of linearly polarized
waves, e.g. in the $x$ direction, with $A^0=0$ and ${\bf A}=(A,0,0)$, provided
that: $i)$ the field $A$ does not depend on the variable $x$ (a {\em transversality}
condition) and $ii)$ $A(t,y,z)$ satisfies the single equation:

\begin{equation}
\partial_\mu\partial^\mu A + \xi\epsilon_0 c^2\left(\partial_\mu\partial^\mu A\partial_\nu A\partial^\nu A+
2\partial_\mu A\partial_\nu^\mu A\partial^\nu A\right)=0,
\label{covariant}
\end{equation}
where we have used the convention $g_{\mu\nu}={\rm diag}(1,-1,-1,-1)$ for the metric tensor.
In non-relativistic notation, Eq. (\ref{covariant}) can be written as
\begin{eqnarray}
\label{non_lin_polarized}
&0=\partial_y^2 A+\partial_z^2 A-\partial_t^2 A   + \xi\epsilon_0 c^2 \left\lbrace\right.&\\ \nonumber
&\left[ \right.\left.(\partial_t A)^2-3(\partial_y A)^2 -(\partial_z A)^2 \right]\partial_y^2A  + &\\ \nonumber
&\left[\right. \left. (\partial_t A)^2 - (\partial_y A)^2 - 3 (\partial_z A)^2\right] \partial_z^2  A  - &\\ \nonumber
&\left[\right.  \left. 3(\partial_t A)^2 - (\partial_y A)^2 - (\partial_z A)^2 \right] \partial_t^2 A  + &\\ \nonumber
&4\left( \partial_z A  \partial_t A \partial_z\partial_t A- \partial_z A  \partial_y A
\partial_z\partial_y A  +  \partial_y A \partial_t A \partial_y\partial_t A\right) \left. \right\rbrace ,
&\nonumber
\end{eqnarray}
where $\partial_y \equiv \frac{\partial }{\partial y}$,
$\partial_z \equiv \frac{\partial }{\partial z}$ and $\partial_t
\equiv \frac{\partial }{c \partial t}$.

Hereafter, we will restrict our discussion to this case of linearly-polarized solution.
The orthogonality relation ${\bf E}\cdot {\bf B}=0$ is then automatically satisfied,
and the effective Lagrangian Eq. (\ref{L}) reduces to
${\cal L}={\cal L}_0\left(1+ \xi{\cal L}_0\right)$. Note that the plane-wave
solutions of the linear Maxwell equations,
such as ${\cal A}\cos\left({\bf k}\cdot{\bf r}- \omega t\right)$,
where ${\cal A}$ is a constant, ${\bf k}=(0,k_y,k_z)$ and $\omega=c \vert {\bf k}\vert$,
are still solutions of Eq. (\ref{covariant}). However, we expect
that the non-linear terms proportional to $\xi$, due to the QED correction,
will spoil the superposition principle.

\section{A numerical solution of the full nonlinear equation (symmetric configuration)}

In general, the numerical solution of a nonlinear wave equation such as
Eq. (\ref{non_lin_polarized}) is a formidable problem. In fact,
it is of an higher order (and is much more complicated) than the
nonlinear Shr\"odinger equation in three dimensions, which is nevertheless
a highly nontrivial and rich system\cite{nlse}. There are two main difficulties in dealing
with Eq. (\ref{non_lin_polarized}). First of all, the direct numerical integration
is practically impossible when the integrating interval is much larger than the wavelength.
Second, one has to find convenient boundary conditions.

In this section, we will study a particular configuration for which
a numerical solution of the full nonlinear equation Eq. (\ref{non_lin_polarized}) can be found.
Let us first consider two counter-propagating plane waves that travel along the $z$-axis,
for simplicity having the same phase at the space-time origin.
The corresponding analytical solution of the linear wave equation (that can be obtained
by setting $\xi=0$ in Eq. (\ref{non_lin_polarized})) would be

\begin{eqnarray}
\label{counterprop_lin}
A_{\rm lin}(t,z)& = &\frac{\cal A}{2}\left[\cos(k z- \omega t)+\cos(k z+
\omega t)\right]\\ \nonumber
& = & {\cal A}\cos(\omega t)\cos(k z) ,
\end{eqnarray}
where ${\cal A}$ is a constant amplitude, ${\bf k}=(0,0,k)$ is the wave vector,
and $\omega=c k$ is the angular frequency. It is easy to see that Eq. (\ref{counterprop_lin})
can also be considered as the analytical solution of the linear wave equation
satisfying the boundary conditions

\begin{eqnarray}
\label{boundary_con}
A(t,0)& = &{\cal A}\cos(\omega t),\\ \nonumber
A\left(-\frac{\pi}{\omega},z\right)& = &A\left(\frac{\pi}{\omega},z\right),\\ \nonumber
\partial_z A(t,0)& = &0 ,
\end{eqnarray}
where for convenience we chose as the integration interval a `small'
cuboid of time-dimension $2 \pi/\omega$ and space-dimension $2 \pi/k$.
With such a choice, the linear wave equation can also be integrated numerically,
obtaining a result that coincides (within the numerical error) with the analytical one.

Of course, in the case of the linear equation this result is trivial. However, we will
use it here as a guide in order to find a solution of the nonlinear Eq. (\ref{non_lin_polarized}).
In fact, although we do not know any analytical solution of the latter equation,
we can find its numerical solution that satisfies the same boundary conditions of Eq. (\ref{boundary_con})
using the same small cuboid as the integration interval.

\begin{figure}[htb]
{\centering \resizebox*{1\columnwidth}{!}{\includegraphics{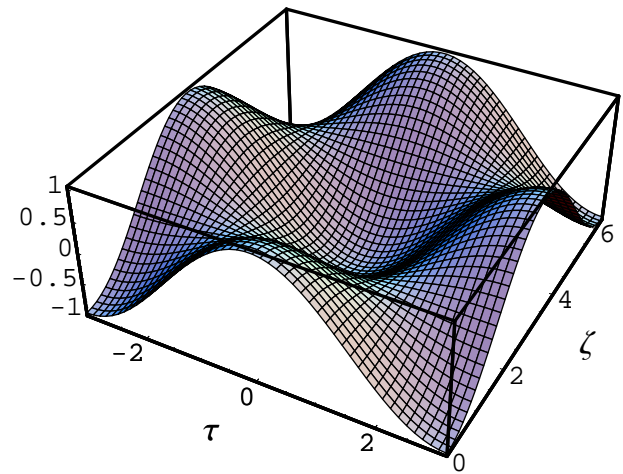}} \par}
\caption{Numerical solution $A_{\rm num}/{\cal A}$ of Eqs. (\ref{non_lin_polarized}) and (\ref{boundary_con})
for $\xi \epsilon_0 {\cal A}^2 \omega^2=0.01$, as a function of the 
adimensional time and space coordinates $\tau\equiv\omega t$ and $\zeta\equiv k z$.}
\label{fig1}
\end{figure}

\begin{figure}[htb]
{\centering \resizebox*{1\columnwidth}{!}{\includegraphics{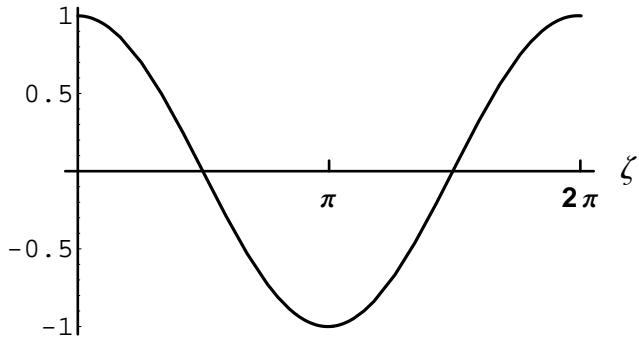}} \par}
\caption{Zero-time plot of $A_{\rm num}/{\cal A}$, as a function of the 
adimensional space coordinate $\zeta\equiv k z$.}
\label{fig2}
\end{figure}

\begin{figure}[htb]
{\centering \resizebox*{1\columnwidth}{!}{\includegraphics{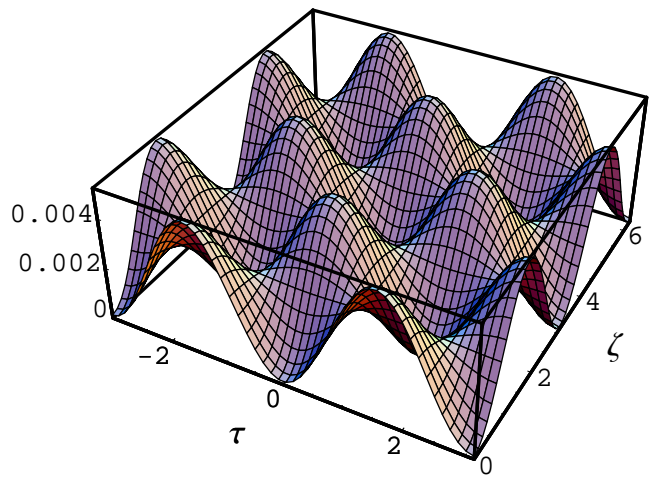}} \par}
\caption{Energy density corresponding to the solution 
$A_{\rm num}$, as a function of the 
adimensional time and space coordinates $\tau\equiv\omega t$ and $\zeta\equiv k z$.}
\label{fig3}
\end{figure}

With these conditions, the result of the numerical integration of Eq. (\ref{non_lin_polarized})
is shown in Figs. \ref{fig1} and \ref{fig2} for a choice of parameters 
such that $\xi \epsilon_0 {\cal A}^2 \omega^2=0.01$. The corresponding energy density, 
obtained as

\begin{eqnarray}
\label{denergy}
\rho &=& T_{00}\\ \nonumber
&=& \frac{\epsilon_0}{2}({\bf E}^2+c^2{\bf B}^2)+\frac{\xi}{4}
\epsilon_0^2 ({\bf E}^2-c^2{\bf B}^2)\left(3 {\bf E}^2+c^2{\bf
B}^2\right),
\end{eqnarray}
is plotted in Fig. \ref{fig3}.

As far as we know, this is the first time that a numerical solution of the full nonlinear wave
equation Eq. (\ref{non_lin_polarized}) is obtained. Of course, it corresponds to a very
simple particular case and a short space-time evolution.
However, we will see later that this solution
will allow us to reach very important conclusions and prove the viability of useful
approximation methods.

\begin{figure}[htb]
{\centering \resizebox*{1\columnwidth}{!}{\includegraphics{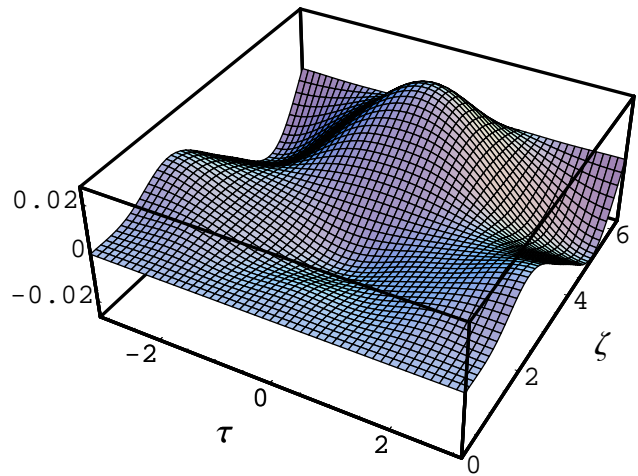}} \par}
\caption{Relative error $\left(A_{\rm num}-A_{\rm lin}\right)/{\cal A}$
of the linear approximation, as a function of the 
adimensional time and space coordinates $\tau\equiv\omega t$ and $\zeta\equiv k z$}.
\label{fig4}
\end{figure}

For the moment, we note that our numerical solution, as given in
Figs. \ref{fig1}, \ref{fig2} and \ref{fig3}, shows an oscillatory behaviour. 
However, the time-average of the energy density turns out to be approximately constant
along the $z$-evolution,
giving $\xi\bar \rho=\frac{\omega}{2\pi}\xi\int_{-\pi/\omega}^{\pi/\omega}\rho dt=0.00252$. 
As we have discusses above, this value of the product $\xi\rho$ 
is several order of magnitude larger than
what can be achieved in the laboratory in the next decades. 
Of course, we will use realistic values of $\rho$ when we will 
present our proposals of experiments in the last sections.
For the moment, it is interesting to note that even such 
an enormous energy density is still small enough so that the 
effect of the nonlinear terms gives a small
correction to the linear evolution in the short distance.
In fact, we see from Fig. \ref{fig4} that the relative difference between the linear and the
nonlinear evolution, in our integration interval which is of the order of the wavelength,
is of the order of few percent, i.e. of the same order than the adimensional 
parameter $\xi \epsilon_0 {\cal A}^2 \omega^2=0.01$. 
This result is not surprising, and will provide a justification for the 
perturbatively-motivated variational approach that we will use in the next sections.

However, from Fig. \ref{fig4} itself, we can also appreciate that 
the difference between the linear and nonlinear behaviour tends to increase 
along the $z$ evolution, so that it can be expected that it will eventually become large
after a distance much larger than the wavelength. In the next section, we will see an
analytical argument that confirms this expectation.

\begin{figure}[htb]
{\centering \resizebox*{1\columnwidth}{!}{\includegraphics{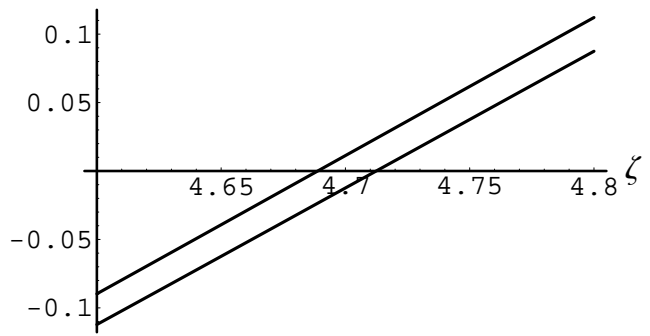}} \par}
\caption{Detail of the zero-time functions $A_{\rm num}/{\cal A}$ 
(upper curve) and $A_{\rm lin}/{\cal A}$ 
(lower curve), for values of the  
adimensional space coordinate $\zeta\equiv k z$ close to the second zero.}
\label{fig5}
\end{figure}

Finally, in Fig. \ref{fig5} we compare the $z$-evolution of the
solutions $A_{\rm num}$ and $A_{\rm lin}$ in a greater detail 
for values of $z$ around the second zero of the solutions.
We see that $A_{\rm num}$ anticipates $A_{\rm lin}$, and the corresponding phase shift
can be evaluated numerically if we define an effective wave vector component $k_z$ 
by computing the value $z_0$ corresponding to $A_{\rm lin}(0,z_0)=0$, and setting $k_z z_0=3 \pi/2$.
The numerical determination of the zero gives $\zeta_0=k z_0=4.68885$, so that $k_z=1.0050 k$.
We will provide a full explanation for this result in the following section.

\section{Variational approximation (symmetric configuration)}

Although it can be considered as an interesting achievement
due to its simplicity and lack of previous approximations,
in practice the numerical solution that has been discussed above
can only be obtained in the special
configuration of two counter-propagating, in-phase waves,
and for short propagation (of the order of the wavelength).
In this section, we will consider the same configuration,
and we will look for a variational approximation that will
provide an {\it analytical}
solution which is valid even for a long evolution.
After showing that this variational solution is
consistent with the numerical one of the previous section,
we will obtain a method to study different configurations that cannot
be integrated numerically in a simple and direct way.

We will thus consider again the two counterpropagating waves,
that would be described by
Eq. (\ref{counterprop_lin}) in the linear case.
Note that any of the two crossing waves, $\frac{\cal A}{2}\cos(k z- \omega t)$
and $\frac{\cal A}{2}\cos(k z+ \omega t)$, when taken alone, would be a
solution of both the linear and non-linear equations, provided that $\omega=c k$.
However, their
superposition would only solve the linear equations of motion.

As we discussed in Ref. \cite{prl99},
in all the experimental configurations that can be studied in the present and the near future,
the product $\xi\rho$ will be so small,
that the non-linear correction will act in a perturbative way,
progressively modifying the form of $A(t,y,z)$ as the wave proceeds along the $z$ direction.
We will therefore make the ansatz
\begin{equation}
A = {\cal A}\left[\alpha(z)\cos(k z) +\beta(z)\sin(k z)\right]\cos(\omega t),
\label{ansatz1}
\end{equation}
allowing for the generation of the other
linearly-independent function $\sin(k z)$
(we will take $\alpha(0)=1$ and $\beta(0)=0$). Note that the invariance of
Eq. (\ref{non_lin_polarized}) under time reversal guarantees that, if the initial
behavior is proportional to $\cos(\omega t)$, which is even under time inversion,
no uneven term proportional to $\sin(\omega t)$ will be generated.

According to the Variational Method, the best choice for the functions
$\alpha(z)$ and $\beta(z)$ corresponds to a local minimum of the
effective action $\Gamma$, after averaging out the time-dependence as follows:

\begin{equation}
\Gamma =\int_{-\infty}^{\infty}d z \left(\frac{ \omega
}{2\pi} \int_{0}^{2\pi/\omega}dt {\cal
L}\right).
\label{Gamma_constraint}
\end{equation}

The expression that is obtained by this procedure is still quite complicated,
due to the presence of the trigonometric functions of multiples of $k z$.
However, as we shall see below, when $\xi\rho\ll 1$ this $z$-dependence is much faster
than that of the envelop functions $\alpha$ and
$\beta$. Therefore, it is a very good approximation to perform the $z$ integral
in two steps: first, we average over a period, treating $\alpha$ and
$\beta$ as if they were constant. In this way, we get rid of the trigonometric
functions. At this point, we allow again the $z$ dependence
of the envelop functions. If we call $\bar\Gamma$ the average action that is obtained
by this procedure, we will then minimize it with respect to the
functions $\alpha$ and $\beta$ by solving the equations $\delta \bar\Gamma/\delta
\alpha=0$ y $\delta\bar\Gamma/\delta \beta=0$.

After a long but straightforward
computation, and neglecting the non-linear terms involving derivatives
of the functions $\alpha$ and $\beta$, since they give a
smaller contribution as we discussed in Ref. \cite{prl99},
these two equations can be written as

\begin{equation}
\frac{\alpha''}{2 k}+\beta' +\chi (\alpha^2+\beta^2) \alpha=0,
\label{eqal}
\end{equation}

and

\begin{equation}
\frac{\beta''}{2 k}-\alpha' +\chi (\alpha^2+\beta^2) \beta=0,
\label{eqbe}
\end{equation}
where $\chi=\xi \epsilon_0{\cal A}^2 c^2 k^3/ 2$ is a parameter that describes the
leading nonlinear effects.

Now, in the perturbative regime in which $\chi$ is small, 
we can also neglect the second derivatives in the previous 
equations\cite{prl99}, so that an (approximate) analytical 
solution for Eqs. (\ref{eqal}) and (\ref{eqbe}) is given by 
$\alpha(z)=\cos(\chi z)$ and $\beta(z)=-\sin(\chi z)$. 
Substituting this result into the variational ansatz (\ref{ansatz1}), 
and using elementary trigonometry, we obtain

\begin{equation}
A={\cal A} \cos(\omega t) \cos[(k+\chi) z].
\label{varsolution1}
\end{equation}

As a result, we find that the phase of the wave is shifted by a term

\begin{equation}
\Delta \phi=\chi z.
\label{phaseshift1}
\end{equation}

Note that the solution of Eq. (\ref{varsolution1}) can also be written as

\begin{equation}
A=\frac{\cal A}{2} \cos[(k+\chi) z-\omega t]+\frac{\cal A}{2} \cos[(k+\chi) z+\omega t],
\label{varsolsinglephases}
\end{equation}
therefore we see that each of two scattering waves is
phase shifted according to Eq. (\ref{phaseshift1}),
due to the crossing with the other wave. 
A similar behavior can also 
be found in a nonlinear medium\cite{nl2},
although the analogy cannot be pushed too 
far, as we have discussed in 
Ref. \cite{prl99}, where we have shown that 
the vacuum does not present the usual AC Kerr effect.

Note also that the result of Eq. (\ref{varsolution1}) can be stated equivalently
by defining a wave vector as $k_z=k+\chi$,
that satisfies a modified dispersion relation, $\omega=c(k_z-\chi)$.

In order to make definite numerical predictions,
it is convenient to express the amplitude ${\cal A}$ in terms of the energy density,
as given by Eq. (\ref{denergy}).
In our perturbative regime, after time and space average,
this gives $\rho\simeq\epsilon_0{\cal A}^2 \omega^2/4$
with a very good approximation. Thus we get $\chi\simeq 2 \xi\rho k
\label{chi_headon}$, so that the phase shift accumulated after a distance 
$\Delta z$ is

\begin{equation}
\Delta \phi\simeq 2 \xi\rho k \Delta z
\label{dphi_headon_sym}
\end{equation}

We see now that the slow varying envelop approximation was justified
as far as $\chi\ll k$.

For the same choice of parameters that was considered in the previous section,
$\xi \rho\simeq\xi \epsilon_0 {\cal A}^2 \omega^2/4=0.0025$, 
in agreement with the numerical simulation. 
Thus we get $\chi=0.005 k$, which is two order of magnitude smaller than $k$.
Our approximations can then be expected to be
reasonably good even for the extremely large of $\rho$ that we have
chosen here. Note also that this corresponds to a value $k_z=k+\chi=1.005 k$,
in agreement with the result that we obtained from the numerical simulation in 
the previous section.

\begin{figure}[htb]
{\centering \resizebox*{1\columnwidth}{!}{\includegraphics{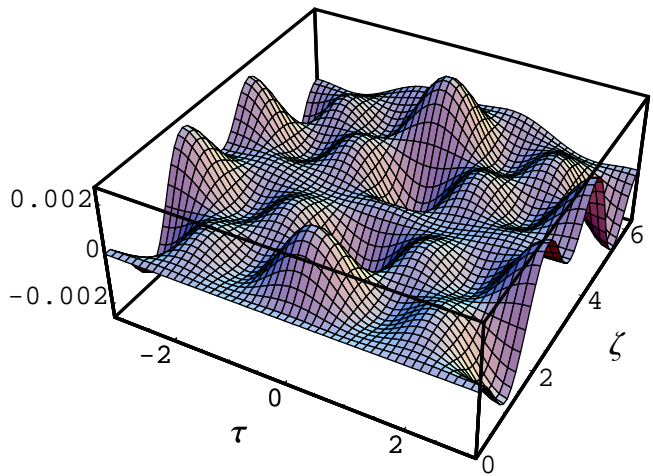}} \par}
\caption{Relative error $\left(A_{\rm num}-A_{\rm var}\right)/{\cal A}$
of the variational approximation, as a function of the 
adimensional time and space coordinates $\tau\equiv\omega t$ and $\zeta\equiv k z$.}
\label{fig6}
\end{figure}

In Fig. \ref{fig6} we compare the corresponding
analytical solution $A_{\rm var}$, given in Eq. (\ref{varsolution1}),
with the numerical solution $A_{\rm num}$ of the QED wave equations
that we have found in the previous section.
Comparing with Fig. \ref{fig4},
we see that the variational solution 
is an order of magnitude closer to the numerical simulation than
the linear evolution, Eq. (\ref{counterprop_lin}), that was obtained
by completely neglecting the nonlinear terms. 
This is a significant improvement in such a short distance.
However, according to the previous discussion,
in the perturbative
regime corresponding to small values of the product $\xi \rho$, 
the variational solution
is expected to be a good approximation even 
when a longer propagation distance is considered along the $z$-axis.
In fact, this expectation is confirmed by Fig.  \ref{fig6} itself,
that shows that the error of the variational solution oscillates 
without any substantial increment in the 
integration interval. 
As we have observed in the previous section, 
this was not the case for the linear solution,
which was incresing its error even in the short distances.
Of course, this is due to the fact that
it does take into account the phase shift, which is the leading effect
due to the nonlinear terms according to our Variational Method.
On the other hand, the agreement of the variational solution with the
numerical simulation can be used as an additional, {\it a posteriori} justification
for our analytical approach. 

\section{Variational approximation (asymmetric configuration)}

In the previous section, we have proved the reliability
of a perturbatively-motivated Variational approach
in the search for approximate solutions of the QED nonlinear wave equation, Eq. (\ref{non_lin_polarized})
In that case, we studied a symmetric configuration in order to compare the result with
the numerical integration of the full equation. With this strong justification in mind,
we can now apply the Variational Method to different configurations, that do not allow
for a direct integration of the full Eq. (\ref{non_lin_polarized}).

In this section, we will study the case of the scattering of a relatively low power
wave with a counterpropagating high power wave, both traveling along the $z$ axis.
In this case, the main effect will be the modification of the low power beam due to the
crossing with the high power one, which would be unaffected in a first approximation.

Using our variational approach, we will describe the high power field by a
wave polarized in the $x$ direction, having the $x$-component
of the vector potential given by

\begin{equation}
A_h={\cal A}\cos(k z + \omega t+\varphi),
\label{A_h}
\end{equation}
where $\varphi$ is an arbitrary phase, describing the unknown phase difference between
the two counterpropagating beams. Such a beam is shot against a low power wave that,
in the absence of the nonlinear QED terms, would be described
by a $x$-component of the vector potential given by $A_l(t,z)=\alpha\cos(k z - \omega t)$,
where the constant $\alpha$ is related to the intensity of the beam.
In other words, the only condition is that the two waves have the same frequency.
Photon-photon scattering will then generate a $z$-dependence of $\alpha$
and an additional term proportional to $\sin(k z - \omega t)$ in the low power wave,
so that

\begin{equation}
A_l(t,z)=\alpha(z)\cos(k z - \omega t)+\beta(z)\sin(k z - \omega t),
\label{A_l}
\end{equation}
with the initial condition $\alpha(0)=\alpha_0$ and $\beta(0)=0$.
Neglecting the effect of the low power beam on the high power wave, we are then lead to
the ansatz

\begin{equation}
A(t,z)=A_h(t,z)+A_l(t,z).
\label{low-high_ansatz}
\end{equation}

According to the Variational Method, we require that the functions
$\alpha(z)$ and $\beta(z)$ correspond to a local minimum of the
effective action $\Gamma$, after averaging out the time-dependence as
described in Eq. (\ref{Gamma_constraint}).
After averaging over the fast dependence on $z$, as discussed
in the previous section, and
neglecting again the non-linear terms involving derivatives
of the functions $\alpha$ and $\beta$, we get the following equations

\begin{equation}
\frac{\alpha''}{2 k}+\beta' +\eta \alpha=0,
\label{eqal_hl}
\end{equation}
and

\begin{equation}
\frac{\beta''}{2 k}-\alpha' +\eta \beta=0,
\label{eqbe_hl}
\end{equation}
where $\eta=2\xi \epsilon_0{\cal A}^2 c^2 k^3$. Note that Eqs. (\ref{eqal_hl}) 
and (\ref{eqbe_hl})
are linear, and they do not show any dependence
of the initial phase difference $\varphi$ between the two beams.
Neglecting the second derivatives, for the same reason that were
discussed in the previous section, we get the solution
$\alpha(z)=\alpha_0 \cos(\eta z)$ and $\beta(z)=-\alpha_0 \sin(\eta z)$.
Note also that this solution is valid for any value of the amplitude
$\alpha_0$ of the low power wave, provided that it is much smaller than
the amplitude of the high power beam,
$\vert\alpha_0\vert\ll \vert{\cal A}\vert$.

As a result, we find that the low power beam becomes

\begin{equation}
A_l(t,z)=\alpha_0\cos[(k+\eta) z - \omega t],
\label{A_l_sol}
\end{equation}

Taking into account that the average energy density of the high power wave
is $\rho=\epsilon_0{\cal A}^2 \omega^2/2$, 
we can compute the phase shift accumulated by the low power wave
after a distance $\Delta z$ as

\begin{equation}
\Delta\Phi=\eta\Delta z\simeq 4 \xi\rho k \Delta z.
\label{phaseshift_lh}
\end{equation}

We stress that this result only depends on the energy density of the high power
wave, as far as it much larger than that of the low power wave. The initial phase
difference is found to be irrelevant.

\section{Proposal of an experiment}

We can now discuss the possibility to test the non-linear properties of the QED vacuum by measuring
small phase changes in one of two crossing laser beams at a very high power laser facility.
In the previous sections, we have studied the
leading QED nonlinear effect on two counterpropagating waves traveling in the
$z$-direction. We have considered two different configurations: a symmetric one,
in which both waves are high power waves with the same initial phase and amplitude;
and an asymmetric configuration, in which only one the two waves is high power,
and they do not need to be in phase. Of course, the second configuration is easier
to be produced experimentally. Moreover, as we shall discuss below,
it leads to a phase shift which is
(at least) twice higher
than that which could be obtained from the symmetric configuration,
for the same experimental facility.

In fact,
the total energy density achievable in the symmetric configuration
is the space-time average due to both waves, that
have to be obtained from the same original high power pulse through a beam
splitter. Therefore, even if we neglect the energy loss when the beams 
split,
the value of $\rho$ in Eq. (\ref{phaseshift_lh}) is roughly the same than that of the single high
power beam of Eq. (\ref{dphi_headon_sym}). Therefore, the asymmetric configuration of section V
produces a phase shift which is roughly twice than that of the symmetric configuration
that we have discussed. 

Therefore, we will propose the following experimental setup, which in
our opinion will provide the simplest and most effective way to look
for optical effects of photon-photon scattering in future exawatt laser facilities.
A common laser pulse is divided in two
beams, A and B, one of which (say A) crosses at a $180^0$ a very high power beam.
As a result, the central
part of the distribution of the beam A has acquired a phase shift $\Delta\Phi$
with respect to beam B, that has propagated freely.
In a experiment corresponding to the parameters of the ELI project in its first
step we have pulses of wavelength $\lambda=800nm$, intensity $I=10^{29}Wm^{-2}$ and
duration $\Delta t=10fs$ which are focused in a spot of diameter $d\approx 10\mu m$.
From Eq. (\ref{phaseshift_lh}), this
results in a phase shift $\Delta\Phi\approx 2\times 10^{-7} rad$ for beam A,
which can be resolved comparing
with the beam B which was not exposed to the effects of QED vacuum. Current
techniques like spectrally resolved two-beam coupling, which can be applied for ultrashort
pulses\cite{kang97}, can be used to this purpose.

It is interesting to note that the sensitivity of this method for the detection of 
photon-photon scattering may be enhanced by a suitable choice of the combination 
of the intensity $I$, the wavelength $\lambda$ and the time duration $\Delta t$ 
that enter in Eq. (\ref{phaseshift_lh}). 
In fact, taking into account that $\Delta z\simeq c\Delta t$, 
Eq. (\ref{phaseshift_lh})
implies that 
the most sensitive experimental configuration will be that having the maximum 
value of the combination $I \Delta t/\lambda$. 

Comparing to other alternatives like x-ray probing of QED birefringence, our
system does not need an extra free electron laser and the power requirements of
the system are only one order of magnitude higher. Moreover, the measurement of
the ellipticity and the polarization rotation angle in birefringence experiments
is not yet possible with current technology. Other techniques like four-wave
mixing processes\cite{4wm} require the crossing of at least three beams, with
the corresponding alignment problems and the rest of the requirements are similar
to our proposal. Moreover, the present result also improves significantly
the one arising for two high-power waves that
we discussed in Ref. \cite{prl99}.

\section{Conclusions}

In this paper, we have studied the nonlinear wave equations that 
describe the electromagnetic field in the vacuum taking into account 
the QED corrections. In particular, we have studied the scattering of 
two waves in a symmetric and an asymmetric configuration.
In the first case, we have found a full numerical solution of
the QED nonlinear wave equations for short-distance
propagation. We have then perfomed a variational analysis 
and found an analytical approximation 
which is in very good agreement with our 
numerical solution, but can also be used for long distance propagation.
We have then studied an aymmmetric 
configuration corresponding to the head-on scattering of a 
ultrahigh power with a low power laser beam 
and argued that it is the most promising configuration
for the search of photon-photon scattering in optics experiments.
In particular, we have shown that our previous requirement of phase coherence between
the two crossing beams can be released.
We have then proposed a very simple experiment that can be performed at
future exawatt laser facilities, such as ELI,
by bombarding a low power laser beam with the exawatt bump.
Photon-photon scattering will then be observed by measuring
the phase shift of the low power laser beam, e.g. by
comparing with a third low power laser beam.
This configuration is simpler, and significantly more sensitive,
than that proposed in our previous paper.
Even in the first step of ELI, we have found that the
resulting phase shift will be at least
$\Delta\Phi\approx 2\times 10^{-7} rad$,
which can be easily measured
with present technology.
Finally, we have discussed how the
experimental parameters can be adjusted
to further improve the sensitivity.

{\em Acknowledgments.-} We thank Miguel \'Angel Garc\'{\i}a-March, Pedro Fern\'andez
de C\'ordoba, G\'erard Mourou and Mario Zacar\'es
for useful discussions. One of the authors (D. T.)
would also like to thank the whole InterTech group at the Universidad Polit\'ecnica de
Valencia for its warm hospitality. 
This work was partially supported by contracts PGIDIT04TIC383001PR
from Xunta de Galicia, ACOMP07/221 from Generalitat Valenciana,
FIS2007-29090-E, FIS2007-62560, FIS2005-01189 and TIN2006-12890 from the Government of Spain.

\end{document}